\documentclass[reqno]{amsart}
\usepackage[dvips]{graphicx}
\usepackage{amsmath,amssymb,amsthm,bm,fancybox,fullpage,dashbox}
\usepackage{bm}
\usepackage[all]{xy}
\usepackage{cite}
\usepackage{verbatim}
\usepackage{enumerate}
\usepackage{url}
\usepackage{color}
\usepackage{tikz-cd, mathdots}

\usepackage{setspace}

\theoremstyle{definition}

\newtheorem{theorem}{Theorem}[section]

\newtheorem{proposition}[theorem]{Proposition}
\newtheorem{definition}[theorem]{Definition}
\newtheorem{remark}[theorem]{Remark}
\newtheorem{example}[theorem]{Example}

\newcommand{\ra}{\rightarrow}

\newcommand{\lra}{\longrightarrow}

\newcommand{\ol}{\overline}
\newcommand{\wt}{\widetilde}

\newcommand{\aut}{\mathsf{Aut}}

\newcommand{\shuffle}{\mathsf{shuffle}}
\newcommand{\F}{\mathcal{F}}

\newlength{\ralen}
\newcommand{\back}{\raisebox{\ralen}{\framebox(11,12){{\large \textbf{?}}}}}

\newcommand{\crd}[1]{\raisebox{\ralen}{\framebox(11,12){#1}}}

\newcommand{\md}[1]{\textcolor{black}{#1}}
\newcommand{\f}[1]{\mathsf{#1}}

\title{Automorphism Shuffles for Graphs and Hypergraphs and Its Applications}
\author{Kazumasa Shinagawa
\and
Kengo Miyamoto}

\address[K. Shinagawa]{Ibaraki University, 4-12-1 Nakanarusawa, Hitachi, Ibaraki, 316-8511, Japan; National Institute of Advanced Industrial Science and Technology (AIST), Tokyo Waterfront Bio-IT Research Building 2-4-7 Aomi, Koto-ku, Tokyo, 135-0064, Japan.}
\email{kazumasa.shinagawa.np92@vc.ibaraki.ac.jp}
\address[K. Miyamoto]{Ibaraki University, 4-12-1 Nakanarusawa, Hitachi, Ibaraki, 316-8511, Japan.}
\email{kengo.miyamoto.uz63@vc.ibaraki.ac.jp}

\begin{document}
\maketitle
\begin{abstract}
In card-based cryptography, a deck of physical cards is used to achieve secure computation. 
A shuffle, which randomly permutes a card-sequence along with some probability distribution, ensures the security of a card-based protocol. 
The authors proposed a new class of shuffles called graph shuffles, which randomly permutes a card-sequence by an automorphism of a directed graph (New Generation Computing 2022). 
For a directed graph $G$ with $n$ vertices and $m$ edges, such a shuffle could be implemented with pile-scramble shuffles with $2(n+m)$ cards. 
In this paper, we study graph shuffles and give an implementation, an application, and a slight generalization of them. 
First, we propose a new protocol for graph shuffles with $2n+m$ cards. 
Second, as a new application of graph shuffles, we show that any cyclic group shuffle, which is a shuffle over a cyclic group, is a graph shuffle associated with some graph. 
Third, we define a hypergraph shuffle, which is a shuffle by an automorphism of a hypergraph, and show that any hypergraph shuffle can also be implemented with pile-scramble shuffles. 
\end{abstract}

\section{Introduction}

\subsection{Background}

\emph{Card-based cryptography} is one of active research areas in cryptography. 
It enables secure computation by using a deck of physical cards. 
In card-based protocols, a deck of physical cards is used to achieve secure computation by hands. 
Thus, it is easy to understand the correctness and the security of protocols, even for non-experts who are unfamiliar with cryptography. 
In fact, there are some reports on an educational application of card-based cryptography for teaching cryptography (e.g., Cornell University \cite{Marcedone15}, University of Waterloo \cite{CHL13}, Tohoku University \cite{MizukiTeaching16}, and a Japanese elementary school \cite{ShinagawaTeaching22}). 

In 2014, Mizuki and Shizuya \cite{MS14} defined a mathematical model of card-based cryptography. 
On the one hand, Mizuki--Shizuya model helps to find some new protocols and to prove some impossibility results (e.g., \cite{KochAC15}). 
On the other hand, it allows some operations which are not clear how to implement physically. 
In particular, the Mizuki--Shizuya model allows the use of shuffles where it is non-trivial to implement physically. 
It is undesirable since card-based cryptography is easy to perform and easy to understand visually. 

Let $\mathfrak{S}_n$ be the symmetric group of degree $n$, $\Pi \subset \mathfrak{S}_n$ a set of permutations, and $\F$ a probability distribution on $\Pi$. 
A \emph{shuffle} $(\shuffle, \Pi, \F)$ for a card-sequence of $n$ cards is an operation to obtain a permuted card-sequence by some $\pi \in \Pi$, where $\pi$ is chosen according to $\F$. 
When we consider $(\shuffle, \Pi, \F)$, we may assume that no player knows which $\pi$ is chosen. 
It is said to be \emph{uniform closed} if $\Pi$ is a group and $\F$ is the uniform distribution. 
\emph{The uniform closed shuffle over $\Pi$} denoted by $(\shuffle, \Pi)$ is a shuffle $(\shuffle, \Pi, \F)$ for the uniform distribution $\F$. 

Given a shuffle $(\shuffle, \Pi, \F)$, it is unclear how to implement it even for uniform closed shuffles. 
In this paper, we study the class of \emph{graph shuffles} (see Subsection \ref{ss:graph}), which is a subclass of uniform closed shuffles including various well-known classes of shuffles such as random cuts \cite{Bo89}, random bisection cuts \cite{MizukiFAW09}, pile-shifting shuffles \cite{ShinagawaProvSec15}, and pile-scramble shuffles \cite{IshikawaUCNC15} and so on. 

\subsection{Related Work}

Koch and Walzer \cite{KW21} showed that any uniform closed shuffle $(\shuffle, \Pi)$ for any group $\Pi$ can be implemented by \emph{random cuts}. 
Although it is worthwhile to show how to implement an arbitrary uniform closed shuffle, it requires many operations (at least the size of the group $\Pi$). 
On the other hand, our protocol requires a small number of operations. 

Saito, Miyahara, Abe, Mizuki, and Shizuya \cite{SaitoTPNC20} showed that every shuffle $(\shuffle, \Pi, \F)$ can be implemented by pile-shifting shuffles. 
Although it is worthwhile to show that every shuffle can be implemented theoretically, it requires many cards (at least $n\cdot |\Pi|$ cards). 
On the other hand, our protocol requires a relatively small number of cards. 

\subsection{Contribution}

\begin{table}[t]
\begin{center}
\caption{Summary of our results and the previous work}
\label{table:summary}
\begin{tabular}{|c|c|c|} \hline
& \# of cards & \# of shuffles \\ \hline
\multicolumn{3}{l}{$\circ$\, \textbf{Graph shuffle protocol (directed, $n$ vertices, and $m$ edges)}} \\ \hline
Miyamoto--Shinagawa \cite{MiyamotoNGC22} & $2(n+m)$ & $d+1$ \\ \hline
Subsection \ref{ss:graphprotocol} & $2n+m$ & $n+2k$ \\ \hline
\multicolumn{3}{l}{$\circ$\, \textbf{Graph shuffle protocol (undirected, $n$ vertices, and $m$ edges)}} \\ \hline
Miyamoto--Shinagawa \cite{MiyamotoNGC22} & $2(n+2m)$ & $d+1$ \\ \hline
Subsection \ref{ss:graphprotocol} & $2(n+m)$ & $n+2d$ \\ \hline
Subsection \ref{ss:hypergtaphs protocol} & $n+2m$ & $n+d+1$ \\ \hline
\multicolumn{3}{l}{$\circ$\, \textbf{Hypergraph shuffle protocol ($n$ vertices and $m$ hyperedges)}} \\ \hline
Subsection \ref{ss:hypergtaphs protocol} & $n+\sum_{i=1}^m|e_i|$ & $n+d'+\ell$ \\ \hline
\end{tabular}
\end{center}
\centerline{$d$ is the number of distinct degrees, $k$ is the number of distinct outdegrees, }
\centerline{$d'$ is the number of distinct degrees, and $\ell$ is the number of distinct sizes of hyperedges. }
\end{table}

In this paper, we focus on the class of graph shuffles introduced by Miyamoto and Shinagawa \cite{MiyamotoNGC22}. 
According to \cite{MiyamotoNGC22}, a graph shuffle associated with a directed graph with $n$ vertices and $m$ edges can be implemented with pile-scramble shuffles and $2(n+m)$ cards.
Our first main result is that we show that it can be implemented with pile-scramble shuffles and $2n+m$ cards.
Thus, the number of cards is more efficient than that of \cite{MiyamotoNGC22}. 
We remark that the number of shuffles 
Second, as an application of graph shuffles, we show that every cyclic group shuffle, which is a uniform closed shuffle over a cyclic subgroup of $\mathfrak{S}_n$, is a graph shuffle. 
It yields that every cyclic group shuffle can be implemented with pile-scramble shuffles only. 
In the last, we consider hypergraphs. 
A hypergraph is a generalization of undirected graphs whose edge (called a hyperedge) is a subset of vertices rather than a pair of vertices. 
We introduce a \emph{hypergraph shuffle}, which is a uniform closed shuffle over the automorphism group of a hypergraph, and design a hypergraph shuffle protocol for any hypergraph $G$. 
It requires $n + \sum_{i=1}^m|e_i|$ cards, where $n$ is the number of vertices, and $m$ is the number of hyperedges. 
Since an undirected graph is a hypergraph, our hypergraph shuffle protocol implies a graph shuffle protocol for undirected graphs. 
As a result, one can reduce the number of cards. 
Our results are summarized in Table \ref{table:summary}. 


\section {Preliminaries}

\subsection {Pile-scramble shuffles}

A \emph{pile-scramble shuffle} is a uniform closed shuffle that randomly permutes $n$ piles of $m$ cards. 
Suppose that we have the following card-sequence:
\[
\underbrace{\underset{\f{x}_{1,1}}{\back} \, \cdots \, \underset{\f{x}_{1,m}}{\back}}_{\f{pile}[1]} \, 
\underbrace{\underset{\f{x}_{2,1}}{\back} \, \cdots \, \underset{\f{x}_{2,m}}{\back}}_{\f{pile}[2]} \, 
\cdots \, 
\underbrace{\underset{\f{x}_{n,1}}{\back} \,\cdots \, \underset{\f{x}_{n,m}}{\back}}_{\f{pile}[n]} ~.
\]
An $(n,m)$-pile-scramble shuffle $\mathsf{PSS}_{(n,m)}$ is a uniform closed shuffle that transforms the above card-sequence into the following card-sequence:
\[
\underbrace{\underset{\f{x}_{\alpha_1,1}}{\back} \, \cdots \, \underset{\f{x}_{\alpha_1,m}}{\back}}_{\f{pile}[\alpha_1]} \, 
\underbrace{\underset{\f{x}_{\alpha_2,1}}{\back} \, \cdots \, \underset{\f{x}_{\alpha_2,m}}{\back}}_{\f{pile}[\alpha_2]} \, 
\cdots \, 
\underbrace{\underset{\f{x}_{\alpha_n,1}}{\back} \, \cdots \, \underset{\f{x}_{\alpha_n,m}}{\back}}_{\f{pile}[\alpha_n]} ~,
\]
where $(\alpha_1, \alpha_2, \ldots, \alpha_n)$ is $\pi(1, 2, \ldots, n)$ for some $\pi \in S_n$ chosen uniformly at random. 

\subsection {Generalized Pile-scramble Protocol}

Suppose that we have $n$ piles where the $i$-th pile consists of $n_i$ cards. 
For a positive integer $s$, we put $A^{(s)} = \{i \mid n_i = s\}$. 
Suppose that $(A^{(s_1)}, A^{(s_2)}, \ldots, A^{(s_t)})$ is a partition of $\{1, 2, \ldots, n\}$ and $A^{(s_j)} = \{i_{j,1}, i_{j,2}, \ldots, i_{j,k}\}$ for $k = |A^{(s_j)}|$. 
For each $1 \leq j \leq t$, we apply a pile-scramble shuffle $\mathsf{PSS}_{(k,s_j)}$ to the card-sequence consisting of the $i_{j,1}, i_{j,2}, \ldots, i_{j,k}$-th piles. 
We define \emph{a generalized pile-scramble protocol} by the above procedure. 

\begin{example}
Suppose that we have the following card-sequence:
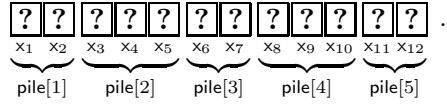
\[
\underbrace{\underset{\f{x}_1}{\back} \, \underset{\f{x}_2}{\back}}_{\f{pile}[1]} \, 
\underbrace{\underset{\f{x}_3}{\back} \, \underset{\f{x}_4}{\back} \, \underset{\f{x}_5}{\back}}_{\f{pile}[2]} \, 
\underbrace{\underset{\f{x}_6}{\back} \, \underset{\f{x}_7}{\back}}_{\f{pile}[3]} \, 
\underbrace{\underset{\f{x}_8}{\back} \, \underset{\f{x}_9}{\back} \, \underset{\f{x}_{10}}{\back}}_{\f{pile}[4]} \, 
\underbrace{\underset{\f{x}_{11}}{\back} \, \underset{\f{x}_{12}}{\back}}_{\f{pile}[5]} ~.
\]
A generalized pile-scramble protocol for the above card-sequence is a sequence of pile-scramble shuffles: a pile-scramble shuffle to $(\f{pile}[1],\f{pile}[3], \f{pile}[5])$ and a pile-scramble shuffle to $(\f{pile}[2], \f{pile}[4])$. 
Applying it, we have one of the following $12$ sequences:
\[
\begin{cases}
(\f{pile}[1], \f{pile}[2], \f{pile}[3], \f{pile}[4], \f{pile}[5]),\\
(\f{pile}[1], \f{pile}[2], \f{pile}[5], \f{pile}[4], \f{pile}[3]),\\
(\f{pile}[3], \f{pile}[2], \f{pile}[5], \f{pile}[4], \f{pile}[1]),\\
(\f{pile}[3], \f{pile}[2], \f{pile}[1], \f{pile}[4], \f{pile}[5]),\\
(\f{pile}[5], \f{pile}[2], \f{pile}[1], \f{pile}[4], \f{pile}[3]),\\
(\f{pile}[5], \f{pile}[2], \f{pile}[3], \f{pile}[4], \f{pile}[1]),\\
(\f{pile}[1], \f{pile}[4], \f{pile}[3], \f{pile}[2], \f{pile}[5]),\\
(\f{pile}[1], \f{pile}[4], \f{pile}[5], \f{pile}[2], \f{pile}[3]),\\
(\f{pile}[3], \f{pile}[4], \f{pile}[5], \f{pile}[2], \f{pile}[1]),\\
(\f{pile}[3], \f{pile}[4], \f{pile}[1], \f{pile}[2], \f{pile}[5]),\\
(\f{pile}[5], \f{pile}[4], \f{pile}[1], \f{pile}[2], \f{pile}[3]),\\
(\f{pile}[5], \f{pile}[4], \f{pile}[3], \f{pile}[2], \f{pile}[1]).
\end{cases}
\]
\; 
\end{example} 

\subsection{Graph Shuffles}\label{ss:graph}

First, we recall some fundamentals from graph theory; for details, see \cite{CZ}.

\emph{A directed graph} is a quadruple $G=(V_G, E_G, s_G, t_G)$ consisting of two sets: $V_G$ (whose elements are called vertices) and $E_G$ (whose elements are called directed edges), and two maps $s_G,t_G: E_G\to V_G$ which associates to its \emph{source} $s_G(\alpha)\in V_G$ and its \emph{target} $t(\alpha) \in V_G$ for $\alpha\in E_G$, respectively. 
A directed edge $\alpha$ with source $a$ and target $b$ is usually denoted by $a\xrightarrow{\alpha}b$. 
A directed graph $G=(V_G, E_G,s_G,t_G)$ is said to be \emph{finite} if both $V_G$ and $E_G$ are finite sets. In this paper, we deal with finite directed graphs. 
For a vertex $v\in V_G$, we define two sets $v^+$ and $v^-$ by
$v^+=\{\alpha\in E_G\mid s_G(\alpha)= v\}$ and $v^-=\{\alpha\in E_G\mid t_G(\alpha)= v\}$.
The cardinality of $v^+$ and $v^-$, denoted by $\mathsf{out}(v)$ and $\mathsf{in}(v)$, are called \emph{the outdegree} and \emph{the indegree}, respectively.
We remark that any undirected graph is regarded as a directed graph by changing each undirected edge to a 2-cycle
$\begin{xy}
                     (0,0)*[o]+{\bullet}="1",(10,0)*[o]+{\bullet}="2",
                     \ar @<1mm> "1";"2"^{}
                     \ar @<1mm>"2";"1"^{}
            \end{xy}.$

Let $G$ and $G'$ be two directed graphs.
A pair of maps $f=(f_0,f_1)$ consisting of $f_0:V_G\to V_{G'}$ and $f_1:E_G\to E_{G'}$ is a \emph{morphism of directed graphs} if it satisfies the equation $(f_0\times f_0)\circ(s_G\times t_G)=(s_{G'}\times t_{G'})\circ f_1$. In addition, if $f_0$ and $f_1$ are bijective, $f$ is called \emph{an isomorphism} between $G$ and $G'$. 
In particular, an isomorphism between $G$ and itself is called \emph{an automorphism} of $G$.
We denote by $\mathsf{Iso}(G,G')$ the set of all isomorphisms between $G$ and $G'$, and we write $\mathsf{Iso}_0(G,G')$ for all bijection $f_0$ such that $f=(f_0,f_1)\in \mathsf{Iso}(G,G')$ for some $f_1$.
When $G=G'$, we set $\mathsf{Aut}(G)=\mathsf{Iso}(G,G)$, and $\mathsf{Aut}_0(G)=\mathsf{Iso}_0(G,G)$. 
The set $\mathsf{Aut}(G)$ has the group structure by using the composition of maps as a product and is called \emph{the automorphism group} of $G$. 
It is easy to check that the group structure of  $\mathsf{Aut}(G)$ induces a group structure of $\mathsf{Aut}_0(G)$. 
Then, we regard $\mathsf{Aut}_0(G)$ as a subgroup of $\mathfrak{S}_{|V_G|}$.

Now, we recall the definition of graph shuffles \cite{MiyamotoNGC22}.

\begin{definition}
Let $G$ be a directed graph. 
The \emph{graph shuffle} associated with $G$ is the uniform closed shuffle over $\aut_0(G)$. 
\end{definition}

\subsection{Graph Shuffle Protocol}\label{ss:graphprotocol}

Let $G$ be a directed graph with $n$ vertices. 
A \emph{graph shuffle protocol} for $(\shuffle, \mathsf{Aut}_0(G))$ is a card-based protocol that implements a graph shuffle associated with $G$. 
Given a card-sequence $\f{x}$ of $n$ cards as an input sequence,  it outputs a card-sequence $\f{y} = \sigma (\f{x})$ for $\sigma \in \aut_0(G)$ as follows:
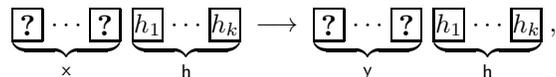
\[
\underbrace{\back \, \cdots \, \back}_{\f{x}} ~ \underbrace{\crd{$h_1$}\, \cdots \, \crd{$h_k$}}_{\f{h}} ~ \lra~ \underbrace{\back \, \cdots \, \back}_{\f{y}} ~ \underbrace{\crd{$h_1$}\, \cdots \, \crd{$h_k$}}_{\f{h}}~,
\]
where $\f{h}$ is a card-sequence of helping cards. 
It is said to be \emph{correct} if the chosen automorphism $\sigma \in \aut_0(G)$ is distributed uniformly at random. 
It is said to be \emph{secure} if a probability distribution of the chosen automorphism $\sigma \in \aut_0(G)$ and a probability distribution of the visible sequence trace (see \cite{MS14} for the definition) of the protocol are stochastically independent. 

\subsection{Miyamoto--Shinagawa's Graph Shuffle Protocol}\label{ss:MiyamotoShinagawa}

Let $G$ be a directed graph with $n$ vertices.
Miyamoto--Shinagawa's graph shuffle protocol \cite{MiyamotoNGC22} for $(\shuffle, \mathsf{Aut}_0(G))$  requires two kinds of cards, black-cards $\crd{1}~\crd{2}~\cdots~\crd{$n$}$ and red-cards $\crd{$\ol{1}$}~\crd{$\ol{2}$}~\cdots~\crd{$\ol{n}$}$~.
It proceeds as follows:
\begin{enumerate}
\item[(1)] Let $\mathsf{x}$ be an input card-sequence.
Place the cards as follows:
\[ 
\underbrace{\back \, \cdots \, \back}_{\f{x}} \,  
\underbrace{\back \, \cdots \, \back}_{\f{pile}[1]} \, \underbrace{\back \, \cdots \, \back}_{\f{pile}[2]} \, \cdots \, \underbrace{\back \, \cdots \, \back}_{\f{pile}[n]}~,
\]
where $\f{pile}[i]$ ($1 \leq i \leq n$) is defined by 
\[
\f{pile}[i]= \underset{\ol{i}}{\back}\, \underbrace{\underset{i}{\back} \, \underset{i}{\back} \, \cdots \, \underset{i}{\back}}_{\mathsf{in}(i)+\mathsf{out}(i) \text{ cards}}.
\]


\item[(2)] Apply a generalized pile-scramble protocol to $n$ piles $(\f{pile}[1], \f{pile}[2], \ldots, \f{pile}[n])$. 
Then we obtain a card-sequence:
\[
\underbrace{\back \, \cdots \, \back}_{\f{x}} ~ \underbrace{\back \, \cdots \, \back}_{\f{pile}[\alpha_1]}~ \underbrace{\back\, \cdots \, \back}_{\f{pile}[\alpha_2]}~ \cdots~ \underbrace{\back \, \cdots \, \back}_{\f{pile}[\alpha_n]}~.
\]
Here, $(\alpha_1, \alpha_2, \ldots, \alpha_n)$ is a permutation of $(1, 2, \ldots, n)$ given by the generalized pile-scramble protocol. 

\item[(3)] For each vertex $i\in V_G$, we define $\f{vertex}[i]$ by 
\[
\f{vertex}[i]= \underset{\ol{\alpha_i}}{\back}\, \underset{\f{x}_i}{\back}~.
\]
For each edge $i \ra j \in E_G$, we define $\f{\md{edge}}[i \ra j]$ by 
\[
\f{\md{edge}}[i \ra j]= \underset{\alpha_i}{\back}\, \underset{\alpha_j}{\back}~.
\]
Place $n+m$ piles as follows:
\[
\underbrace{\back\,\back}_{\f{vertex}[1]}\,\cdots\,\underbrace{\back\,\back}_{\f{vertex}[n]}~~
\underbrace{\back\,\back}_{\f{\md{edge}}[e_1]}\,\cdots\,\underbrace{\back\,\back}_{\f{\md{edge}}[e_m]},
\]
where $E_G = \{e_1, e_2, \ldots, e_m\}$. 

\item[(4)] Apply $\f{PSS}_{(m+n,2)}$ to the card-sequence. 
\item[(5)] Open the left card of all piles. 
If it is a black-card, turn over the right card. 
Then sort $n+m$ piles so that the left card lied as $\crd{$\ol{1}$} \, \cdots \, \crd{$\ol{n}$}\, \crd{$1$}\, \cdots \,\crd{$n$}\,$. 
Suppose that we have a card-sequence as follows:
\[
\crd{$\ol{1}$}\,\back~~
\crd{$\ol{2}$}\,\back~~
\cdots~~
\crd{$\ol{n}$}\,\back~~
\crd{$i_1$}\,\crd{$j_1$}~~
\crd{$i_2$}\,\crd{$j_2$}~~
\cdots~~
\crd{$i_m$}\,\crd{$j_m$}\,,
\]

\item[(6)] Define a graph $G'$ by 
\begin{align*}
&V_{G'} = V_G,\\
&E_{G'} = \{i_1\ra j_1, i_2\ra j_2, \ldots, i_m\ra j_m\}.
\end{align*}

\item[(7)] Choose an isomorphism $\psi:G'\to G$. 
Set $\beta_i := \psi^{-1}_0(i)$. 
Output $\f{y} =(\f{y}_1, \f{y}_2, \ldots, \f{y}_n)$, where $\f{y}_i$ is the right next card of $\crd{$\ol{\beta_i}$}\,$. 
\end{enumerate}

\section{Our Graph Shuffle Protocol}\label{s:graph}

In this section, we propose a new graph shuffle protocol for an arbitrary directed graph, which is a more efficient protocol than that of \cite{MiyamotoNGC22} in terms of the number of cards. 

\subsection{Our Idea}

The protocol in \cite{MiyamotoNGC22} needs $n + m$ piles of two cards: $n$ piles for vertices and $m$ piles for edges. 
Each pile $\f{vertex}[i]$ corresponding to a vertex $i$ consists of a card representing a (randomized) vertex $\ol{\alpha_i}$ and the $i$-th input card $\f{x}_i$. 
Each pile $\f{\md{edge}}[i \ra j]$ corresponding to an edge consists of cards representing a randomized edge $(\ol{\alpha_i}, \ol{\alpha_j})$. 
Thus, the number of cards is $2(n+m)$. 

Our idea to reduce the number of cards is that we make \emph{$n$ piles only}. 
For each vertex, we assign a pile as follows. 
A pile associated with the $i$-th vertex consists of the $i$-th input card $\f{x}_i$, a card representing a (randomized) vertex $\ol{\alpha_i}$, and $\f{out}(i)$ cards representing (randomized) outgoing edges from $i$. 
Thus, the number of cards is $2n+\sum_{i \in V_G}\f{out}(i) = 2n+m$. 

\subsection{Our Protocol}\label{ss:graphprotocol}

Let $G = (V_G,E_G,s_G,t_G)$ be an arbitrary directed graph with $n$ vertices and $m$ edges. 
We set $V_G=\{1,2,\ldots, n\}$. 
Let $\f{x} = (\f{x}_1,\f{x}_2,\ldots, \f{x}_n)$ be an input card-sequence for the shuffle. 
Our protocol proceeds as follows. 
\begin{enumerate}
\item[(1)] Place the cards as follows:
\[ 
\underbrace{\back \, \cdots \, \back}_{\f{x}} \,  
\underbrace{\crd{1} \, \cdots \, \crd{1}}_{\f{pile}[1]} \, \underbrace{\crd{2} \, \cdots \, \crd{2}}_{\f{pile}[2]} \, \cdots \, \underbrace{\crd{$n$} \, \cdots \,  \crd{$n$}}_{\f{pile}[n]}~,
\]
where $\f{pile}[i]$ ($1 \leq i \leq n$) is a pile of cards consists of $\f{in}(i)+1$ copies of $\crd{$i$}\,$. 

\item[(2)] Apply a generalized pile-scramble protocol to $n$ piles $(\f{pile}[1], \f{pile}[2], \ldots, \f{pile}[n])$. 
Then we obtain a card-sequence:
\[
\underbrace{\back \, \cdots \, \back}_{\f{x}} ~ \underbrace{\back \, \cdots \, \back}_{\f{pile}[\alpha_1]}~ \underbrace{\back\, \cdots \, \back}_{\f{pile}[\alpha_2]}~ \cdots~ \underbrace{\back \, \cdots \, \back}_{\f{pile}[\alpha_n]}~.
\]
Here, $(\alpha_1, \alpha_2, \ldots, \alpha_n)$ is a permutation of $(1, 2, \ldots, n)$ given by the generalized pile-scramble protocol. 

\item[(3)] Suppose $\{t_G(\alpha) \mid \alpha \in i^+\} =  \{w_{i,1}, w_{i,2}, \ldots, w_{i,\f{out}(i)}\}$ as a multiset for each $i\in V_G$. 
Let $\f{vertex}[i]$ be a pile of cards defined as follows:
\[
\f{vertex}[i] = \underset{\f{x}_i}{\back}\,\underset{\alpha_i}{\back} \, \underset{\alpha_{w_{i,1}}}{\back} \, \underset{\alpha_{w_{i,2}}}{\back} \, \cdots \, \underset{\alpha_{w_{i,\f{out}(i)}}}{\back}
\]
Place the card as follows:
\[ 
\underbrace{\back \, \back \, \cdots \, \back}_{\f{vertex}[1]} \, \underbrace{\back \, \back \, \cdots \, \back}_{\f{vertex}[2]} \, \cdots \, \underbrace{\back \, \back \, \cdots \, \back}_{\f{vertex}[n]}
\]

\item[(4)] For each $\f{vertex}[i]$, except the first and second cards, apply $\f{PSS}_{(\mathsf{out}(i), 1)}$ to the $\mathsf{out}(i)$ cards. 
Let $\f{vertex}'[i]$ be the resultant pile. 
Then, the current card-sequence is
\[ 
\underbrace{\back \, \back \, \cdots \, \back}_{\f{vertex}'[1]} \, \underbrace{\back \, \back \, \cdots \, \back}_{\f{vertex}'[2]} \, \cdots \, \underbrace{\back \, \back \, \cdots \, \back}_{\f{vertex}'[n]}~.
\]

\item[(5)] Apply a generalized pile-scramble protocol to $n$ piles $(\f{vertex}'[1], \f{vertex}'[2], \ldots, \f{vertex}'[n])$. 
Then we obtain the following card-sequence:
\[
\underbrace{\back \, \back \, \cdots \, \back}_{\f{vertex}'[\beta_1]} \, \underbrace{\back \, \back \, \cdots \, \back}_{\f{vertex}'[\beta_2]} \, \cdots \, \underbrace{\back \, \back \, \cdots \, \back}_{\f{vertex}'[\beta_n]}~.
\]
Here, $(\beta_1, \beta_2, \ldots, \beta_n)$ is a permutation of $(1, 2, \ldots, n)$ given by the generalized pile-scramble protocol. 

\item[(6)] For each pile, turn over all cards except the first card. 
Then sort $n$ piles so that the second card is in ascending order as follows:
\[
\underbrace{\back \, \crd{1} \, \cdots \cdots }_{\f{vertex}'[\gamma_1]} \, 
\underbrace{\back \,\crd{2} \, \cdots\cdots }_{\f{vertex}'[\gamma_2]} \, 
\cdots \, 
\underbrace{\back \,\crd{$n$} \, \cdots \cdots }_{\f{vertex}'[\gamma_n]}~.
\]
Here, $(\gamma_1, \gamma_2, \ldots, \gamma_n)$ is a permutation of $(\beta_1, \beta_2, \ldots, \beta_n)$. 
Let $\f{y}_i$ be the first card of $\f{vertex}'[\gamma_i]$. 
For each $i$, we suppose that 
\[
\f{vertex}'[\gamma_i] = \underset{\f{y}_i}{\back} \, \crd{$i$} \, \crd{\footnotesize{$j^i_1$}} \, \crd{\footnotesize{$j^i_2$}} \, \cdots\, \crd{\footnotesize{$\,j^i_{d_i}$}}
\]

\item[(7)] We define a graph $G'$ by
\begin{itemize}
\item $V_{G'} = V_G$,
\item $E_{G'} = \displaystyle \bigcup_{i = 1}^n \{ i \ra j^{i}_s \mid 1 \leq s \leq d_i \}$.
\end{itemize}

\item[(8)] Take an isomorphism $\psi:G'\to G$. 
Set $\f{z} = (\f{z}_1, \f{z}_2, \ldots, \f{z}_n)$ with $\f{z}_i = \f{y}_{\psi^{-1}_0(i)}$. 
Arrange the card-sequence as follows:
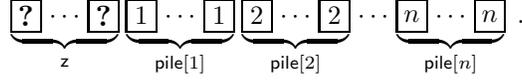
\[
\underbrace{\back \, \cdots \, \back}_{\f{z}} \,  
\underbrace{\crd{1} \, \cdots \, \crd{1}}_{\f{pile}[1]} \, \underbrace{\crd{2} \, \cdots \, \crd{2}}_{\f{pile}[2]} \, \cdots \, \underbrace{\crd{$n$} \, \cdots \,  \crd{$n$}}_{\f{pile}[n]}~.
\]
The output card-sequence for the input $\f{x}$ is $\f{z}$. 
\end{enumerate}

\begin{remark}
For the number of cards, the above protocol requires $2n+m$ cards. 
For the number of shuffles, it requires $2k+n$ PSSs, where $k = |\{\f{out}(i) \mid i \in V_G\}|$. 
Note that it is not necessary to apply the PSSs to a card-sequence in Step (4) if $\f{out}(i)=1$.
Thus, if we set $n'=|\{i \mid \f{out}(i)\geq 2\}|$, the protocol requires exactly $2k+n'$ PSSs.
\end{remark}

\subsection{Correctness and Security}\label{ss:correctness}

For an arbitrary directed graph $G$ with $V_G=\{1,2,\ldots, n\}$, we set
\begin{align*}
&H_G^{\f{in}} = \{\pi \in \mathfrak{S}_n \mid \f{in}(i) = \f{in}(\pi(i))~\text{for all $1 \leq i \leq n$}\},\\
&H_G^{\f{out}}= \{\pi \in \mathfrak{S}_n \mid \f{out}(i) = \f{out}(\pi(i))~\text{for all $1 \leq i \leq n$}\}.
\end{align*}
Then $H_G^{\f{in}}$ and $H_G^{\f{out}}$ are subgroups of $\mathfrak{S}_n$, and $\aut_0(G)$ is a subgroup of $H_G^{\f{in}}$ and $H_G^{\f{out}}$ since every automorphism preserves the indegree and outdegree of each vertex. 

Let $\f{x}=(\f{x}_1, \f{x}_2, \ldots, \f{x}_n)$ be an input card-sequence, $\f{y}=(\f{y}_1, \f{y}_2, \ldots, \f{y}_n)$ the card-sequence described in Step (8), and $\f{z}=(\f{z}_1, \f{z}_2, \ldots, \f{z}_n)$ the corresponding output card-sequence. 
We take permutations $\sigma, \tau \in \mathfrak{S}_n$ such that $\sigma^{-1}(i) = \alpha_i$ in Step (2) and $\tau^{-1}(i) = \beta_i$ in Step (5), respectively. 
One can easily check that $\sigma \in H_G^{\f{in}}$ and $\tau \in H_G^{\f{out}}$. 
Let $G'$ be the directed graph defined at Step (7). 
Let $\chi = (\chi_0,\chi_1) \in \mathsf{Iso}(G,G')$ be an isomorphism such that $\chi_0 = \sigma^{-1}$ and $\psi = (\psi_0, \psi_1) \in \mathsf{Iso}(G',G)$ an isomorphism chosen at Step (8). 


Since $\aut_0(G)$ is a subgroup of $H_G^{\f{in}} \cap H_G^{\f{out}}$, we have $\sigma = \pi \sigma'$ and $\tau = \tau'\pi'$ for some $\sigma' \in H_G^{\f{in}}$, $\tau' \in H_G^{\f{out}}$, and $\pi, \pi' \in \aut_0(G)$. 
Then we have 
\begin{equation}\label{eq1}
\tau \sigma = \tau'\pi' \pi \sigma' = \tau'\pi'' \sigma'
\end{equation}
 for $\pi'' = \pi' \pi \in \aut_0(G)$. 
It implies from $\psi_0 = \wt{\pi} \sigma'$ for some $\wt{\pi} \in \aut_0(G)$ that 
\begin{equation}\label{eq2}
\f{z} = \psi_0 \sigma^{-1}(\f{x}) =  \wt{\pi} \sigma' (\pi \sigma')^{-1}(\f{x}) = \wt{\pi} \pi^{-1}(\f{x}).
\end{equation}

As the distribution of $\pi'$ is uniform over $\aut_0(G)$, we conclude that the distributions of $\wt{\pi}$ and $\pi''$ are independent by the above equations (\ref{eq1}) and (\ref{eq2}). 
It yields that $\wt{\pi} \pi^{-1}$ is uniform over $\aut_0(G)$. 
Therefore, our protocol is secure and correct. 

\subsection{Example Execution of Our Protocol}\label{ss:execution}

We show an execution of our protocol for the following graph $G$: 
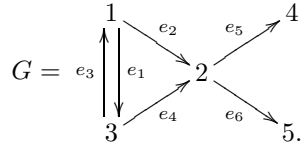
\[ G=  \begin{xy}
                     (0,8)*[o]+{1}="1",(0,-8)*[o]+{3}="2",(12,0)*[o]+{2}="3",(24,8)*[o]+{4}="4",(24,-8)*[o]+{5.}="5",
                     \ar @<1mm>"1";"2"^{e_1}
                     \ar @<1mm>"2";"1"^{e_3}
                     \ar "1";"3"^{e_2}
                     \ar "2";"3"_{e_4}
                     \ar "3";"4"^{e_5}
                     \ar "3";"5"_{e_6}
            \end{xy} \]
Then, the automorphism group is 
\[
\mathsf{Aut}_0(G) = \{ \mathsf{id}, (1\;3), (4\;5), (1\;3)(4\;5) \}.
\]
Let $\f{x} = (\f{x}_1,\f{x}_2,\f{x}_3,\f{x}_4,\f{x}_5)$ be an input card-sequence. 

\begin{enumerate}
\item[(1)] Place the cards as follows:
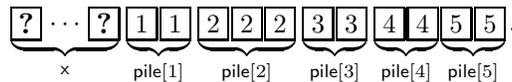
\[ 
\underbrace{\back \, \cdots \, \back}_{\f{x}} \,  
\underbrace{\crd{1} \, \crd{1}}_{\f{pile}[1]} \, \underbrace{\crd{2} \, \crd{2} \, \crd{2}}_{\f{pile}[2]} \, \underbrace{\crd{3} \, \crd{3}}_{\f{pile}[3]} \, \underbrace{\crd{4} \, \crd{4}}_{\f{pile}[4]}\underbrace{\crd{5} \, \crd{5}}_{\f{pile}[5]}.
\]

\item[(2)] Apply a generalized pile-scramble protocol to five piles $(\f{pile}[1], \f{pile}[2], \f{pile}[3], \f{pile}[4], \f{pile}[5])$. 
Then we obtain a card-sequence:
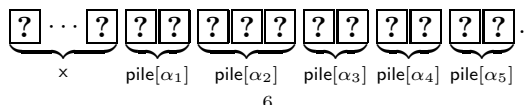
\[ 
\underbrace{\back \, \cdots \, \back}_{\f{x}} \,  
\underbrace{\back \, \back}_{\f{pile}[\alpha_1]} \, 
\underbrace{\back \, \back \, \back}_{\f{pile}[\alpha_2]} \, 
\underbrace{\back \, \back}_{\f{pile}[\alpha_3]} \, 
\underbrace{\back \, \back}_{\f{pile}[\alpha_4]} \, 
\underbrace{\back \, \back}_{\f{pile}[\alpha_5]}.
\]
Here, $(\alpha_1, \alpha_2, \alpha_3, \alpha_4, \alpha_5)$ is a random permutation of $(1,2,3,4,5)$ with $\alpha_2 = 2$. 

\item[(3)] Arrange the piles of cards as follows:
\begin{align*}
& \f{vertex}[1] = \underset{\f{x}_1}{\back}\,\underset{\alpha_1}{\back} \, \underset{\alpha_{2}}{\back} \, \underset{\alpha_{3}}{\back}~, 
~ \f{vertex}[4] = \underset{\f{x}_4}{\back}\,\underset{\alpha_4}{\back}~,\\
& \f{vertex}[2] = \underset{\f{x}_2}{\back}\,\underset{\alpha_2}{\back} \, \underset{\alpha_{4}}{\back} \, \underset{\alpha_{5}}{\back}~,
~\f{vertex}[5] = \underset{\f{x}_5}{\back}\,\underset{\alpha_5}{\back}~,\\
& \f{vertex}[3] = \underset{\f{x}_3}{\back}\,\underset{\alpha_3}{\back} \, \underset{\alpha_{1}}{\back} \, \underset{\alpha_{2}}{\back}~.
\end{align*}

\item[(4)] Apply PSS to each $\f{vertex}[i]$ except the first and second cards. 
\begin{align*}
& \f{vertex'}[1] = \underset{\f{x}_1}{\back}\,\underset{\alpha_1}{\back} \, \underset{\alpha_{2}'}{\back} \, \underset{\alpha_{3}'}{\back}\ , \quad \alpha_2',\alpha_3'\in\{\alpha_2,\alpha_3\}.\\
& \f{vertex'}[2] = \underset{\f{x}_2}{\back}\,\underset{\alpha_2}{\back} \, \underset{\alpha_{4}'}{\back} \, \underset{\alpha_{5}'}{\back}\ , \quad \alpha_4',\alpha_5'\in\{\alpha_4,\alpha_5\}.\\\
& \f{vertex'}[3] = \underset{\f{x}_3}{\back}\,\underset{\alpha_3}{\back} \, \underset{\alpha_{1}'}{\back} \, \underset{\alpha_{2}''}{\back}\ , \quad \alpha_1',\alpha_2''\in\{\alpha_1,\alpha_2\}.\\\
& \f{vertex'}[4] = \underset{\f{x}_4}{\back}\,\underset{\alpha_4}{\back}~,  \quad \f{vertex'}[5] = \underset{\f{x}_5}{\back}\,\underset{\alpha_5}{\back}~.
\end{align*}

\item[(5)] Apply a generalized pile-scramble protocol to five piles $(\f{vertex}'[1], \f{vertex}'[2], \f{vertex}'[3], \f{vertex}'[4], \f{vertex}'[5])$. 
Then we obtain a card-sequence:
\[
\underbrace{\back \, \back \, \back \, \back}_{\f{vertex}'[\beta_1]} \, 
\underbrace{\back \, \back \, \back \, \back}_{\f{vertex}'[\beta_2]} \, 
\underbrace{\back \, \back \, \back \, \back}_{\f{vertex}'[\beta_3]} \, 
\underbrace{\back \, \back}_{\f{vertex}'[\beta_4]} \, 
\underbrace{\back \, \back}_{\f{vertex}'[\beta_5]}.
\]
Here, $(\beta_1, \beta_2, \beta_3)$ is a random permutation of $(1,2,3)$ and $(\beta_4, \beta_5)$ is a random permutation of $(4, 5)$. 

\item[(6)] For each pile, turn over all cards except the first card. 
Suppose that we obtain a card-sequence as follows:
\[ 
\underbrace{{\back} \, \crd{2} \, \crd{5} \, \crd{3}}_{\f{vertex}'[\beta_1]} \, \underbrace{{\back}\, \crd{1} \, \crd{2} \, \crd{4}}_{\f{vertex}'[\beta_2]} \, \underbrace{{\back} \, \crd{4} \, \crd{2} \, \crd{1}}_{\f{vertex}'[\beta_3]}\, \underbrace{{\back} \, \crd{5}}_{\f{vertex}'[\beta_4]}\, \underbrace{{\back} \, \crd{3}}_{\f{vertex}'[\beta_5]}.
\]
Then sort five piles so that the second card is in ascending order as follows:
\[ 
 \underbrace{{\back}\, \crd{1} \, \crd{2} \, \crd{4}}_{\f{vertex}'[\beta_2]}  \, 
\underbrace{{\back} \, \crd{2} \, \crd{5} \, \crd{3}}_{\f{vertex}'[\beta_1]} \, 
\underbrace{{\back} \, \crd{3}}_{\f{vertex}'[\beta_5]}\, 
\underbrace{{\back} \, \crd{4} \, \crd{2} \, \crd{1}}_{\f{vertex}'[\beta_3]}\, 
\underbrace{{\back} \, \crd{5}}_{\f{vertex}'[\beta_4]}.
\]
Set $(\gamma_1, \gamma_2, \gamma_3, \gamma_4, \gamma_5) = (\beta_2, \beta_1, \beta_5, \beta_3, \beta_4)$. 
Let $\f{y}_i$ be the first card in $\f{vertex}'[\gamma_i]$. 

\item[(7)] From the opened symbols, the graph $G'$ is defined as follows:
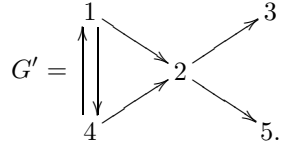
\[ G'=  \begin{xy}
                     (0,8)*[o]+{1}="1",(0,-8)*[o]+{4}="2",(12,0)*[o]+{2}="3",(24,8)*[o]+{3}="4",(24,-8)*[o]+{5.}="5",
                     \ar @<1mm>"1";"2"^{}
                     \ar @<1mm>"2";"1"^{}
                     \ar "1";"3"^{}
                     \ar "2";"3"_{}
                     \ar "3";"4"^{}
                     \ar "3";"5"_{}
            \end{xy} \]
\item[(8)] Take an isomorphism $\psi: G'\to G$. 
For example, $\psi_0(1)=3, \psi_0(2)=2, \psi_0(3)=4, \psi_0(4)=1, \psi_0(5)=5$. 
We have $(\psi^{-1}_0(1), \psi^{-1}_0(2), \psi^{-1}_0(3), \psi^{-1}_0(4), \psi^{-1}_0(5))=(4,2,1,3,5)$.
The output card-sequence is as follows:
\[
\underset{\f{y}_4}{\back}\ \underset{\f{y}_2}{\back}\ \underset{\f{y}_1}{\back}\ \underset{\f{y}_3}{\back}\ \underset{\f{y}_5}{\back}
\]
\end{enumerate}

\section{Cyclic Group Shuffles via Graph Shuffles}

In this section, we state that every ``cyclic group shuffle'' is a graph shuffle associated with some graph. 

Fix a positive integer $n$.
Let $C$ be an arbitrary cyclic subgroup of $\mathfrak{S}_n$.
The uniform closed shuffle over $C$ is called the \emph{$C$-cyclic group shuffle}.
For instance, the $\langle (1\ 2\ 3\ 4\ 5\ 6)\rangle$-cyclic group shuffle is a random cut for six cards. 
We note that every cyclic group shuffle is not only a random cut. 
For instance, the $\langle (1\ 2)(3\ 4\ 5\ 6)\rangle$-cyclic group shuffle is different from a random cut. 
Indeed, if we apply the $\langle (1\ 2)(3\ 4\ 5\ 6)\rangle$-cyclic group shuffle to $\f{x}=\crd{1}\,\crd{2}\,\crd{3}\,\crd{4}\,\crd{5}\,\crd{6}$ then we have the following card-sequence $\sigma (\mathsf{x})$ with probability $\dfrac{1}{4}$:
\[ \begin{array}{ll}
 \bullet\quad  \crd{1}\ \crd{2}\ \crd{3}\ \crd{4}\ \crd{5}\ \crd{6}  & \text{$\sigma=\mathsf{id}$,}\\
\bullet\quad  \crd{2}\ \crd{1}\ \crd{6}\ \crd{3}\ \crd{4}\ \crd{5}  & \text{$\sigma=(1\ 2)(3\ 4\ 5\ 6)$,}\\
 \bullet\quad \crd{1}\ \crd{2}\ \crd{5}\ \crd{6}\ \crd{3}\ \crd{4}  & \text{$\sigma=(3\ 4\ 5\ 6)^2$,}\\
  \bullet\quad \crd{2}\ \crd{1}\ \crd{4}\ \crd{5}\ \crd{6}\ \crd{3}  & \text{$\sigma=(1\ 2)(3\ 4\ 5\ 6)^3 $.}\\
\end{array} \]

Let $C$ be a subgroup of $\mathfrak{S}_n$ with a generator $g$. 
Assume that a cycle decomposition of $g$ is given as $g=g_1g_2\cdots g_t$, where each $g_i$ is of the form $g_i= (g^{(i)}_0,\ldots, g^{(i)}_{\ell_i-1})$.
Without loss of generality, we may assume that $\ell_i\leq \ell _j$ for any $1\leq i<j\leq t$.
For each cycle $g_i$, we denote by $C^{[i]}$ the following directed cycle:
\[ C^{[i]}=\begin{xy}
                     (0,-2)*[o]+{g^{(i)}_0}="1",(15,-2)*[o]+{g^{(i)}_1}="2",(30,-2)*[o]+{\cdots}="3",(45,-2)*[o]+{g^{(i)}_{\ell_i-2}}="4",(60,-2)*[o]+{g^{(i)}_{\ell_i-1}.}="n",
                     \ar "1";"2"^{}
                     \ar "2";"3"^{}
                     \ar "3";"4"_{}
                     \ar "4";"n"_{}
                     \ar @(lu,ur)"n";"1"_{}
            \end{xy}\]
For distinguish two positive integers $k$ and $k'$, we write $d(k,k')$ for the greatest common divisor of 
$\ell_k$ and $\ell_{k'}$.
Now, we define a directed graph $Q(g)$ as follows.
\begin{itemize}
\item $V_{Q(g)}= \{1, 2, \ldots, n\}$,
\item $E_{Q(g)}=\displaystyle  \bigcup_{k=0}^t E_{C^{[k]}}\ \cup \displaystyle  \bigcup_{d(k,k')\neq 1, k < k'} \{g^{(k)}_u\to g^{(k')}_v\mid  u\equiv v\  \mathrm{mod}\ d(k,k')\}$.\\
Here, we sum over all pairs $(k,k')$ such that $d(k,k')\neq 1$ and $k < k'$. 
\end{itemize}
We call the directed graph $Q(g)$ \emph{the gear graph of $g$}.

The following proposition holds from the construction of the gear graph $Q(g)$. 
\begin{proposition}
Let $g\in\mathfrak{S_n}$. 
Then the graph shuffle associated with $Q(g)$ is the $\langle g\rangle$-cyclic group shuffle. 
\end{proposition}

\begin{example} 
The gear graph of 
\[ g=(1\ 2\ 3)(4\ 5\ 6\ 7)(8\ 9\ 10\ 11\ 12\ 13)\in \mathfrak{S}_{13} \]
is given as follows.
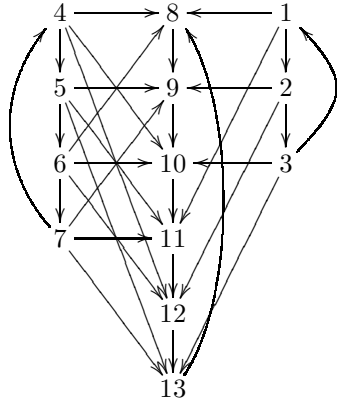
\[\begin{xy}
 (30,0)*[o]+{1}="1",(30,-10)*[o]+{2}="2",(30,-20)*[o]+{3}="3",
 (0,0)*[o]+{4}="4", (0,-10)*[o]+{5}="5",(0,-20)*[o]+{6}="6",(0,-30)*[o]+{7}="7", 
 (15,0)*[o]+{8}="8", (15,-10)*[o]+{9}="9",(15,-20)*[o]+{10}="10", (15,-30)*[o]+{11}="11", (15,-40)*[o]+{12}="12",(15,-50)*[o]+{13}="13",
 \ar "1";"2"^{}
\ar "2";"3"_{}

\ar "4";"5"_{}
\ar "5";"6"_{}
\ar "6";"7"_{}

\ar "8";"9"_{}
\ar "9";"10"_{}
\ar "10";"11"_{}
\ar "11";"12"_{}
\ar "12";"13"_{}

 \ar "4";"8"^{}
  \ar "4";"10"^{}
   \ar "4";"12"^{}
    \ar "5";"9"^{}
  \ar "5";"11"^{}
   \ar "5";"13"^{}
    \ar "6";"8"^{}
\ar "6";"10"^{}
 \ar "6";"12"^{}
   \ar "7";"9"^{}
    \ar "7";"11"^{}
        \ar "7";"13"^{}
        
         \ar "1";"8"^{}
  \ar "1";"11"^{}
   \ar "2";"9"^{}
    \ar "2";"12"^{}
  \ar "3";"10"^{}
   \ar "3";"13"^{}
   
\ar @(ul,dl)  "7";"4"^{}
\ar @(ur,dr)  "3";"1"^{}
\ar @(ur,dr)  "13";"8"^{}
            \end{xy}\]
Therefore, the $\langle g\rangle$-cyclic group shuffle is the graph shuffle associated with $Q(g)$. 
\end{example}

\section{Hypergraph Shuffle Protocol}

In this section, we define a ``hypergraph shuffle'', which is a generalization of (undirected-)graph shuffles, and propose a hypergraph shuffle protocol for an arbitrary hypergraph. 

\subsection{Hypergraph Shuffles}\label{ss:hypergtaphs}

\emph{A hypergraph} is a pair $H=(V_H, \mathcal{E}_H)$ consisting of a set of \emph{vertices} $V_H$ and a family of subsets of $V_H$.
Each element $e^H\in\mathcal{E}_H$ is called \emph{a hyperedge}. 
Note that every undirected graph is a hypergraph such that any hyperedges has exactly two vertices.
In this section, we deal with finite hypergraphs, i.e., the numbers of vertices and hyperedges are finite.
For two hypergraphs $H=(V_H, \mathcal{E}_H)$ and $H'=(V_{H'}, \mathcal{E}_{H'})$, where $\mathcal{E}_H=\{e^H_1,\ldots, e^H_n\}$ and $\mathcal{E}_{H'}=\{e^{H'}_1,\ldots, e^{H'}_n\}$, we say that $H$ and $H'$ are \emph{isomorphic} if there exist a bijective map $f:V_H\to V_{H'}$ and a permutation $\sigma\in\mathfrak{S}_{|E_H|}$ such that, for any $e^H_i\in \mathcal{E}_H$, $f(e^H_i)=e^{H'}_{\sigma(i)}$ holds.
Such a map $f$ is called \emph{an isomorphism} of hypergraphs between $H$ and $H'$. 
We denote by $\mathsf{Iso}(H, H')$ the set of all isomorphisms of hypergraphs between $H$ and $H'$. 
We set $\mathsf{Aut}(H)=\mathsf{Iso}(H,H')$ whose elements are called \emph{automorphisms} of $H$.
By the definition, $\mathsf{Aut}(H)$ is a subgroup of $\mathfrak{S}_{|V_H|}$.

Now we define a class of hypergraph shuffles. 

\begin{definition}
Let $H$ be a hypergraph. 
The \emph{hypergraph shuffle} associated with $H$ is the uniform closed shuffle over $\aut(H)$. 
\end{definition}

\subsection{Our Protocol}\label{ss:hypergtaphs protocol}

Let $H = (V_H,\mathcal{E}_H)$ be any hypergraph having $n$ vertices and $m$ hyperedges. 
We set $V_H=\{1,2,\ldots, n\}$, and $\mathcal{E}_H=\{e_1, e_2, \ldots, e_m\}$. 
Besides $n$ input cards, our protocol requires $\sum_{k=1}^m|e_k|$ helping cards as follows:
\[ 
\underbrace{\crd{$1$} \, \cdots\, \crd{$1$}}_{|e_1|}\, \underbrace{\crd{$2$} \, \cdots\, \crd{$2$}}_{|e_2|}\, \cdots \, \underbrace{\crd{$m$} \, \cdots\, \crd{$m$}}_{|e_m|}\,. \]
Let $\f{x} = (\f{x}_1,\f{x}_2,\ldots, \f{x}_n)$ be an input card-sequence for the shuffle. 
Our protocol proceeds as follows. 

\begin{enumerate}
\item[(1)] Place the cards as follows:
\[ 
\underbrace{\back \, \cdots \, \back}_{\f{x}} \,  
\underbrace{\crd{$1$} \, \cdots \, \crd{$1$}}_{\f{pile}[1]} \, \underbrace{\crd{$2$} \, \cdots \, \crd{$2$}}_{\f{pile}[2]} \, \cdots \, \underbrace{\crd{$m$} \, \cdots \,  \crd{$m$}}_{\f{pile}[m]}~.
\]
Here, $\f{pile}[i]$ ($1 \leq i \leq m$) is a pile of cards consists of $|e_i|$ copies of $\crd{$i$}\,$. 

\item[(2)] Apply a generalized pile-scramble protocol to $m$ piles $(\f{pile}[1], \f{pile}[2], \ldots, \f{pile}[m])$. 
Then we obtain a card-sequence:
\[
\underbrace{\back \, \cdots \, \back}_{\f{x}} ~ \underbrace{\back \, \cdots \, \back}_{\f{pile}[\alpha_1]}~ \underbrace{\back\, \cdots \, \back}_{\f{pile}[\alpha_2]}~ \cdots~ \underbrace{\back \, \cdots \, \back}_{\f{pile}[\alpha_m]}~.
\]
Here, $(\alpha_1, \alpha_2, \ldots, \alpha_n)$ is a permutation of $(1, 2, \ldots, m)$ given by the generalized pile-scramble protocol. 

\item[(3)] For each $i\in V_H$, we set $E_H^{(i)} =  \{j \mid i\in e_j\}$ as a multiset.
We may suppose that $E_H^{(i)}=\{j_1, j_2,\ldots, j_{s_i}\}$ and $j_k\leq j_\ell$ for $k<\ell$.
We then define 
\[
\f{vertex}[i] = \underset{\f{x}_i}{\back}\,\underset{\alpha_{j_1}}{\back} \, \underset{\alpha_{j_2}}{\back} \, \underset{\alpha_{j_3}}{\back} \, \cdots \, \underset{\alpha_{j_{s_i}}}{\back}.
\]
After that, we arrange the cards as follows:
\[ 
\underbrace{\back \, \back \, \cdots \, \back}_{\f{vertex}[1]} \, \underbrace{\back \, \back \, \cdots \, \back}_{\f{vertex}[2]} \, \cdots \, \underbrace{\back \, \back \, \cdots \, \back}_{\f{vertex}[n]}
\]

\item[(4)] For each $i \in V_H$, apply $\f{PSS}_{(|E_H^{(i)}|, 1)}$ to the $|E_H^{(i)}|$ cards appearing in $\f{vertex}[i]$ except for the first card. 
Let $\f{vertex}'[i]$ be the resultant pile. 
We suppose that the current card-sequence is
\[ 
\underbrace{\back \, \back \, \cdots \, \back}_{\f{vertex}'[1]} \, \underbrace{\back \, \back \, \cdots \, \back}_{\f{vertex}'[2]} \, \cdots \, \underbrace{\back \, \back \, \cdots \, \back}_{\f{vertex}'[n]}~.
\]

\item[(5)] Apply a generalized pile-scramble protocol to $n$ piles $(\f{vertex}'[1], \f{vertex}'[2], \ldots, \f{vertex}'[n])$. 
Then we obtain a card-sequence:
\[
\underbrace{\back \, \back \, \cdots \, \back}_{\f{vertex}'[\beta_1]} \, \underbrace{\back \, \back \, \cdots \, \back}_{\f{vertex}'[\beta_2]} \, \cdots \, \underbrace{\back \, \back \, \cdots \, \back}_{\f{vertex}'[\beta_n]}~.
\]
Here, $(\beta_1, \beta_2, \ldots, \beta_n)$ is a permutation of $(1, 2, \ldots, n)$ given by the generalized pile-scramble protocol. 

\item[(6)] For each pile, turn over all cards except the first card. 
We suppose that
\[ \f{vertex}'[\beta_k]=\underset{\mathsf{y}_k}{\back} \, \crd{$j_1$} \, \cdots \, \crd{$j_{\ell_k}$}\quad (k=1,2,\ldots, n). \] 

\item[(7)] We define a hypergraph as follows:
\begin{itemize}
\item $V_{H'} = V_H$,
\item the vartex $k$ belongs to $\widetilde{e_j}$ if and only if \crd{$j$} appears in $\f{vertex}'[\beta_k]$.
We then set $\mathcal{E}_{H'}=\{\widetilde{e_1},\ldots, \widetilde{e_m}\}$.
\end{itemize}

\item[(8)] Take an isomorphism of hypergraphs $\psi:H\to H'$, and set $\f{z} = (\f{z}_1, \f{z}_2, \ldots, \f{z}_n)$ with $\f{z}_i = \f{y}_{\psi(i)}$. 
Arrange the card-sequence as follows:
\[
\underbrace{\back \, \cdots \, \back}_{\f{z}} \,  
\underbrace{\crd{$1$} \, \cdots \, \crd{$1$}}_{\f{pile}[1]} \, \underbrace{\crd{$2$} \, \cdots \, \crd{$2$}}_{\f{pile}[2]} \, \cdots \, \underbrace{\crd{$m$} \, \cdots \,  \crd{$m$}}_{\f{pile}[m]}~.
\]
The output card-sequence for the input $\f{x}$ is $\f{z}$. 
\end{enumerate}


\begin{remark}
For the number of cards, the above protocol requires $n+\sum_{i=1}^m|e_i|$ cards. 
For the number of shuffles, it requires $n + \ell + d'$ PSSs, where $\ell = |\{|e| \mid e \in \mathcal{E}_H\}|$ and $d' = |\{|E_H^{(i)}| \mid i \in V_H\}|$. 
\end{remark}

\begin{remark}
Consider an undirected graph $G$ with $n$ vertices and $m$ undirected edges. 
Let $\vec{G}$ be the corresponding directed graph of $G$ (see Subsection \ref{ss:graph}). 
Since $\aut(G)$ (as a hypergraph) equals to $\aut(\vec{G})$ (as a directed graph), the hypergraph shuffle associated with $G$ and the graph shuffle associated with $\vec{G}$ implement the same shuffle. 
We observe that the hypergraph shuffle protocol requires $n+2m$ cards while the graph shuffle protocol requires $2(n+m)$ cards. 
(Note that $\vec{G}$ has $2m$ directed edges.) 
Thus, the hypergraph shuffle protocol is more efficient than the graph shuffle protocol for undirected graphs in terms of the number of cards. 
\end{remark}

\subsection{Correctness and Security}\label{ss:correctness}

For an arbitrary hypergraph $H = (V_H,\mathcal{E}_H)$ with $V_H=\{1,2,\ldots, n\}$, and $\mathcal{E}_H=\{e_1, e_2, \ldots, e_m\}$, we set
\[
S_H = \{\pi \in \mathfrak{S}_n \mid |E_H^{(i)}| = |E_H^{(\pi(i))}|~\text{for all $1 \leq i \leq n$}\}.
\]
Then $S_H$ is a subgroup of $\mathfrak{S}_n$, and $\aut(H)$ is a subgroup of $P_H$ since every automorphism preserves the number of adjacent edges of each vertex. 

Let $\f{x}=(\f{x}_1, \f{x}_2, \ldots, \f{x}_n)$ be an input card-sequence, $\f{y}=(\f{y}_1, \f{y}_2, \ldots, \f{y}_n)$ the card-sequence described in Step (8), and $\f{z}=(\f{z}_1, \f{z}_2, \ldots, \f{z}_n)$ the corresponding output card-sequence. 
We take permutations $\sigma \in \mathfrak{S}_m$ and $\tau \in \mathfrak{S}_n$ such that $\sigma^{-1}(i) = \alpha_i$ in Step (2) and $\tau^{-1}(i) = \beta_i$ in Step (5), respectively. 
One can easily check that $\tau \in S_H$. 
Let $H'$ be the hypergraph defined at Step (7). 
Let $\psi \in \mathsf{Iso}(H,H')$ by an isomorphism chosen at Step (8). 

Since $\aut(H)$ is a subgroup of $P_H$, we have $\tau = \tau'\pi$ for some $\tau' \in S_H$ and $\pi \in \aut(H)$. 
It implies from $\psi = \tau' \pi'$ for some $\pi' \in \aut(H)$ that 
\begin{equation}\label{eq3}
\f{z} = \psi^{-1} \tau(\f{x}) =  (\tau' \pi')^{-1} \tau'\pi(\f{x}) = (\pi')^{-1} \pi(\f{x}).
\end{equation}

Thanks to the randomization by $\sigma$, the opened symbols in Step (6) reveal nothing about the permutation $\pi$. 
Thus, the choice of $\pi'$ is independent from $\pi$. 
Since the distribution of $\pi$ is uniform over $\aut(H)$, it yields that $(\pi')^{-1} \pi$ is uniform over $\aut(H)$. 
Therefore, our protocol is secure and correct. 

\subsection{Example Execution of Our Protocol}\label{ss:example hypergtaphs protocol}

We show an execution of our protocol for the following hypergraph $H$: 
\begin{itemize}
\item $V_H = \{1,2,3,4,5\}$,
\item $\mathcal{E}_{H}=\{e_1, e_2, e_3\}$, where $e_1 = \{1, 2, 3\}$, $e_2 = \{2, 4\}$, and $e_3 = \{2, 5\}$. 
\end{itemize}
Let $\f{x} = (\f{x}_1,\f{x}_2,\f{x}_3,\f{x}_4,\f{x}_5)$ be an input card-sequence. 

\begin{enumerate}
\item[(1)] Place the cards as follows:
\[ 
\underbrace{\back \, \back \, \back \, \back \, \back}_{\f{x}} \,  
\underbrace{\crd{$1$} \, \crd{$1$} \, \crd{$1$}}_{\f{pile}[1]} \, \underbrace{\crd{$2$} \, \crd{$2$}}_{\f{pile}[2]} \, \underbrace{\crd{$3$} \, \crd{$3$}}_{\f{pile}[3]}~.
\]

\item[(2)] Apply a generalized pile-scramble protocol to $3$ piles $(\f{pile}[1], \f{pile}[2], \f{pile}[3])$. 
Then we obtain a card-sequence:
\[
\underbrace{\back \, \back \, \back \, \back \, \back}_{\f{x}} ~ \underbrace{\back \, \back \, \back}_{\f{pile}[\alpha_1]}~ \underbrace{\back\, \back}_{\f{pile}[\alpha_2]}~ \underbrace{\back\, \back}_{\f{pile}[\alpha_3]}~.
\]
Here, $(\alpha_1, \alpha_2, \alpha_3)$ is a permutation of $(1, 2, 3)$. 

\item[(3)] Arrange the piles of cards as follows:
\begin{align*}
& \f{vertex}[1] = \underset{\f{x}_1}{\back}\,\underset{\alpha_1}{\back}~, 
~\f{vertex}[2] = \underset{\f{x}_2}{\back}\,\underset{\alpha_1}{\back} \, \underset{\alpha_2}{\back} \, \underset{\alpha_3}{\back}~, \\
& \f{vertex}[3] = \underset{\f{x}_3}{\back}\,\underset{\alpha_1}{\back}~,
~ \f{vertex}[4] = \underset{\f{x}_4}{\back}\,\underset{\alpha_2}{\back}~,  \\
& \f{vertex}[5] = \underset{\f{x}_5}{\back}\,\underset{\alpha_3}{\back}~.
\end{align*}

\item[(4)] Apply PSS to the second, third, and fourth cards in $\f{vertex}[2]$. 
Then we have
\begin{align*}
& \f{vertex'}[i] = \f{vertex}[i]\ , \quad i\in\{1,3,4,5\},\\
& \f{vertex'}[2] = \underset{\f{x}_2}{\back}\,\underset{\alpha_{1}'}{\back} \, \underset{\alpha_{2}'}{\back} \, \underset{\alpha_{3}'}{\back}\ , ~\alpha_1',\alpha_2',\alpha_3'\in\{\alpha_1,\alpha_2,\alpha_3\}~.
\end{align*}

\item[(5)] Apply a generalized pile-scramble protocol to $5$ piles $(\f{vertex}'[1], \f{vertex}'[2], \f{vertex}'[3], \f{vertex}'[4], \f{vertex}'[5])$. 
Then we obtain a card-sequence:
\[
\underbrace{\back \, \back}_{\f{vertex}'[\beta_1]} \, \underbrace{\back \, \back \, \back \, \back}_{\f{vertex}'[\beta_2]} \, \underbrace{\back \, \back}_{\f{vertex}'[\beta_3]} \, \underbrace{\back \, \back}_{\f{vertex}'[\beta_4]} \, \underbrace{\back \, \back}_{\f{vertex}'[\beta_5]}~.
\]
Here, $(\beta_1, \beta_2, \beta_3, \beta_4, \beta_5)$ is a permutation of $(1, 2, 3,4,5)$ with $\beta_2 = \alpha_2$. 

\item[(6)] For each pile, turn over all cards except the first card. 
Suppose that we obtain a card-sequence as follows:
\[ 
\underbrace{{\back} \, \crd{$2$}}_{\f{vertex}'[\beta_1]} \, \underbrace{{\back}\, \crd{$3$} \, \crd{$2$} \, \crd{$1$}}_{\f{vertex}'[\beta_2]} \, \underbrace{{\back} \, \crd{$1$}}_{\f{vertex}'[\beta_3]}\, \underbrace{{\back} \, \crd{$3$}}_{\f{vertex}'[\beta_4]}\, \underbrace{{\back} \, \crd{$2$}}_{\f{vertex}'[\beta_5]}.
\]
Let $\f{y}_i$ be the first card in $\f{vertex}'[\beta_i]$. 

\item[(7)] From the opened symbols, the hypergraph $H'$ is defined as follows:
\begin{itemize}
\item $V_{H'} = V_H$,
\item $\mathcal{E}_{H'}=\{\widetilde{e_1},\widetilde{e_2}, \widetilde{e_3}\}$, where $\widetilde{e_1} = \{2, 3\}$, $\widetilde{e_2} = \{1, 2, 5\}$, and $\widetilde{e_3} = \{2, 4\}$. 
\end{itemize}

\item[(8)] 
Take an isomorphism $\psi: H\to H'$. 
For example, $\psi(1)=1, \psi(2)=2, \psi(3)=5, \psi(4)=3, \psi(5)=4$. 
The output card-sequence is as follows:
\[
\underset{\f{y}_1}{\back}\ \underset{\f{y}_2}{\back}\ \underset{\f{y}_5}{\back}\ \underset{\f{y}_3}{\back}\ \underset{\f{y}_4}{\back}
\]
\end{enumerate}

\begin{remark}
We remark that $\aut(H)$ is just the automorphism group of $G$ described in Subsection \ref{ss:execution}. 
Thus, the above hypergraph shuffle protocol and the graph shuffle protocol described in Subsection \ref{ss:execution} implement the same shuffle. 
For the number of cards, the hypergraph shuffle protocol requires $12$ cards and three PSSs, while the graph shuffle protocol requires $16$ cards and six PSSs. 
\end{remark}

\section*{Funding}
K. Shinagawa was partly supported by JSPS KAKENHI 21K17702. 
K. Miyamoto was partly supported by JSPS KAKENHI 20K14302. 

\bibliographystyle{abbrv}
\bibliography{card}




\end{document}